\documentclass[preprint]{aastex701}
\usepackage{amsmath}
\begin{document}

\title{Possible Quasi-Period Oscillation Signals in the Unique Event of \\GRB 250702DBE/EP250702a?}

\author[orcid=0009-0004-9835-353X]{Fei-Fan Song}
\email{songfeifan@ynao.ac.cn}
\affiliation{Yunnan Observatories, Chinese Academy of Sciences, 650011 Kunming, Yunnan Province, People's Republic of China}
\affiliation{University of Chinese Academy of Sciences, 100049 Beijing, People’s Republic of China}
\author[0000-0002-7077-7195]{Jirong Mao}
\email[show]{jirongmao@mail.ynao.ac.cn}
\affiliation{Yunnan Observatories, Chinese Academy of Sciences, 650011 Kunming, Yunnan Province, People's Republic of China}
\affiliation{Center for Astronomical Mega-Science, Chinese Academy of Sciences, 20A Datun Road, Chaoyang District, 100012 Beijing, People's Republic of China}
\affiliation{Key Laboratory for the Structure and Evolution of Celestial Objects, Chinese Academy of Sciences, 650011 Kunming, People's Republic of China}

\begin{abstract}
GRB 250702DBE was time-consequently triggered by GBM onboard the {\it Fermi} satellite. 
It is uncertain which celestial catalog is suitable for this special ultra-long event to belong to. In this paper, we comprehensively investigate the lightcurves obtained by Fermi-GBM detectors. In the energy band of 8-1000 keV, 
no Quasi-Period Oscillation 
(QPO) signals are found in the lightcurve of the first burst 250702D, a possible QPO signal of 0.046 Hz corresponding to a period of 21.7 s is found in the lightcurve of the second burst 250702B, and a possible QPO signal of 0.024 Hz corresponding to a period of 41.7 s 
is found in the lightcurve of last burst 250702E. 
The significance level of the possible QPO signals is comprehensively examined.
In addition, we examine the spectral properties of the sources. In general, a broken power law is suitable for modeling the spectral data from 8 keV to 40 MeV. 
We qualitatively suggest some kinds of celestial object with the periodic characteristic that might be the progenitors of this unique event.   
\end{abstract}

\keywords{High-energy astrophysics}


\section{Introduction}
Some kinds of celestial object have sudden energy release in a short time period, and multiwavelength observation is important for revealing the physical properties. In particular, time variability of radiation is an interesting topic to classify unknown objects and reveal the temporal properties of the sources. In general, high-energy instruments can provide good time resolution in lightcurves. When a trigger is alerted from a high-energy satellite, lightcurves may be first examined to classify the source. Then, one may preliminarily judge the related physical properties.    

On 2025 July 2, the Fermi Gamma-ray Burst Monitor (GBM) triggered three bursts: GRB 250702D at 13:09:02.03 UT (T$_{0}$)
\citep{2025GCN.40886....1F}, GRB 250702B at 13:56:05 UT (T$'_{0}$)\citep{2025GCN.40883....1F}, and GRB 250702E at 16:21:33 UT (T$''_{0}$) \citep{2025GCN.40890....1F}.
These bursts were named as traditional gamma-ray burst (GRB) convention. Then, it was found that they have very similar localizations and they were identified as the same source by different triggers \citep{2025GCN.40931....1N,2025GCN.40891....1N}.
The photon energy distributions from three bursts are primarily concentrated in the 50-300 keV range, with detectable emission extending beyond 1 MeV \citep{2025GCN.40931....1N}. 

Some other high-energy telescopes and ground-based telescopes performed follow-up observations to GRB 250702DBE. At 02:53:44 UTC on 2 July 2025, the Einstein Probe (EP) Wide-field X-ray Telescope (WXT) detected an X-ray transient in the vicinity of the three burst locations, and the source was named as EP250702a. 
\citep{2025GCN.40906....1C}.
NuSTAR, MAXI and Konus-Wind performed the observations to this event, respectively \citep{2025GCN.41014....1O,2025GCN.40910....1K,2025GCN.40914....1F}.
Observations from HST, VLT, Keck I, and other optical telescopes suggested the very red color of the counterpart \citep{2025GCN.40924....1M,2025GCN.41096....1G,2025GCN.40949....1B,2025GCN.41044....1D}.
This event was also detected by submillimeter and radio telescopes \citep{2025GCN.40983....1A,2025GCN.41054....1A,2025GCN.41059....1A,2025GCN.41061....1T}.
In the very high-energy band, the non-detection results was given by H.E.S.S. \citep{2025GCN.41095....1N}.  

The results of the multiwavelength observation were obtained on GRB 250702DBE/EP250702a. However,   
the nature of this event is quite uncertain. By the investigation mainly from the optical observations, \citet{2025arXiv250714286L} suggested several 
possibilities on the progenitors and physical origins of the event. In this paper, we use the lightcurves of GRB 250702DBE/EP250702a obtained from {\it Fermi}-GBM to perform periodic analysis. 
The results of Quasi-Period Oscillation (QPO) detection are crucial to classify this event and further reveal its physical origin.  

We describe the process of data analysis in Section \ref{data}. The results are presented in Section \ref{results}. Conclusions and discussion are given in Section \ref{discussion}. 

\section{Data Analysis}
\label{data}
We take the Fermi Gamma-ray Burst Monitor (GBM) data for GRB 250702D, GRB 250702B, and GRB 250702E\footnote{https://heasarc.gsfc.nasa.gov/FTP/fermi}. The GBM instrument consists of 12 sodium iodide (NaI) detectors, covering an energy range of 8-1000 keV, and two bismuth germanate (BGO) detectors, sensitive to 200 keV–40 MeV \citep{fermigbm2009ApJ...702..791M}. 
Because this emission of the event mainly falls below 1 MeV, we first focus on the data of the NaI detectors. 
Among the 12 NaI detectors, we select the data from the 
detector with the small viewing angle relative to the event, ensuring optimal photon detection efficiency. We use the data for temporal and spectral analysis. In addition, we examine the data obtained from BGO instrument. The data from the detector in BGO with the small viewing angle 
is used. 
We also use the data for temporal and spectral analysis.

Among GBM data products, time-tagged event (TTE) data have the highest time and energy resolution, with a time resolution of 2 $\mu s$ and an energy resolution of 128 channels.
We employ Fermi GBM data tools \citep{GbmDataTools} to extract the lightcurves 
of the bursts from Time-Tagged Event (TTE) data. The lightcurves are generated in the 8-1000 keV energy band with 256 ms temporal binning. 
To model the background counts, we first define the background intervals for each burst. We reference the T$_{90}$ intervals of each burst from the Fermi GBM catalog and appropriately scale the T$_{90}$ intervals to ensure the bursts are primarily contained within these time intervals, while selecting data outside the scaled T$_{90}$ intervals for background fitting. For GRB 250702D, the background intervals are chosen as [T$_{0}$ - 130 s, T$_{0}$ - 100 s] and [T$_{0}$ + 18 s, T$_{0}$ + 116 s]. For GRB 250702B, the background intervals are selected as [T$'_{0}$ - 137 s, T$'_{0}$ - 57 s] and [T$'_{0}$ + 173 s, T$'_{0}$ + 403 s]. For GRB 250702E, the background intervals are defined as [T$''_{0}$ - 135 s, T$''_{0}$ - 20 s] and [T$''_{0}$ + 171 s, T$''_{0}$ + 362 s]. We simulate the detector background by fitting a polynomial function to the selected background intervals, then subtract the best-fit background model from the raw lightcurves to obtain the background-subtracted lightcurves. The method described above to obtain the lightcurves is traditionally used for the analysis of Fermi-GBM data \citep{2014ApJS..211...13V,2016ApJS..223...28N}.

We perform a temporal examination to the lightcurves of GRB 250702DBE/EP250702a. 
In this study, we employ Fast Fourier Transform (FFT) method \citep{FFT} and Weighted Wavelet Z-transform (WWZ) method \citep{foster1996AJ....112.1709F} to analyze 
the lightcurves of this event. 
The FFT method is specifically suited for evenly sampled data\footnote{It is sometimes mistakenly believed that the Lomb-Scargle periodogram (LSP), 
and other methods developed to analyze data with arbitrary 
sampling, have some benefit for the case of evenly sampled data. 
But in fact, LSP is identical to the ordinary periodogram based 
on the Fourier transform used here.}. It is applied to the uniformly sampled lightcurves to transform the time-domain data into the frequency domain.
Here, we apply zero-padding to the lightcurves for FFT analysis.
It is important to note that this process does not improve the true spectral resolution, which is limited by the observation duration. Instead, it provides an interpolated spectrum with a finer frequency bin size. For a single-tone signal, we want to visually identify a peak clearly and identify the frequency closer to the real isolated one. Thus, we can accurately estimate the amplitude.
The WWZ method utilizes Morlet wavelets to convert time-series data into the time-frequency domain. 
The results obtained by the two methods respectively can be cross-checked.
We focus on QPO searching. The FFT and WWZ methods represent commonly used QPO analysis techniques in astrophysics. 
By the two methods, we analyze the lightcurves, 
in which the background intervals are excluded.

We explain the FFT method as below. The power spectral density (PSD) analysis is constrained to the meaningful frequency range between 1/T (T represents observation duration) and the Nyquist frequency ($f_s/2$, with $f_s$ being the sampling frequency).
In particular, we model the FFT PSD as a combination of red and white noise components $P(f)=af^{b}+c$. 
The optimal parameters are determined through Whittle likelihood maximization \citep{Hubner2022ApJS..259...32H}. 
We take the best-fit noise model, and we can determine the 95\%, 99\% and $3\sigma$ confidence level of the noise model. First, we have generated 20,000 simulated lightcurves based on the noise model. These simulations 
mimic
the temporal characteristics of the observed lightcurve, specifically its time interval size, sampling frequency, and root mean square (RMS) variability. Then, we calculate the PSD for each of these simulated lightcurves. Based on these PSDs, the distribution of power values at each frequency can be derived. For each frequency, we use the distribution to calculate the power thresholds corresponding to the 95\%, 99\%, and $3\sigma$ confidence levels of the noise model. Thus, we can judge a possible QPO at a certain frequency when we see a power value exceed a certain confidence level. This is the way that we judge how often the periodogram is large.  

\section{Results} 
\label{results}
\begin{figure}[h!]
    \centering
    \includegraphics[width=0.75\textwidth]{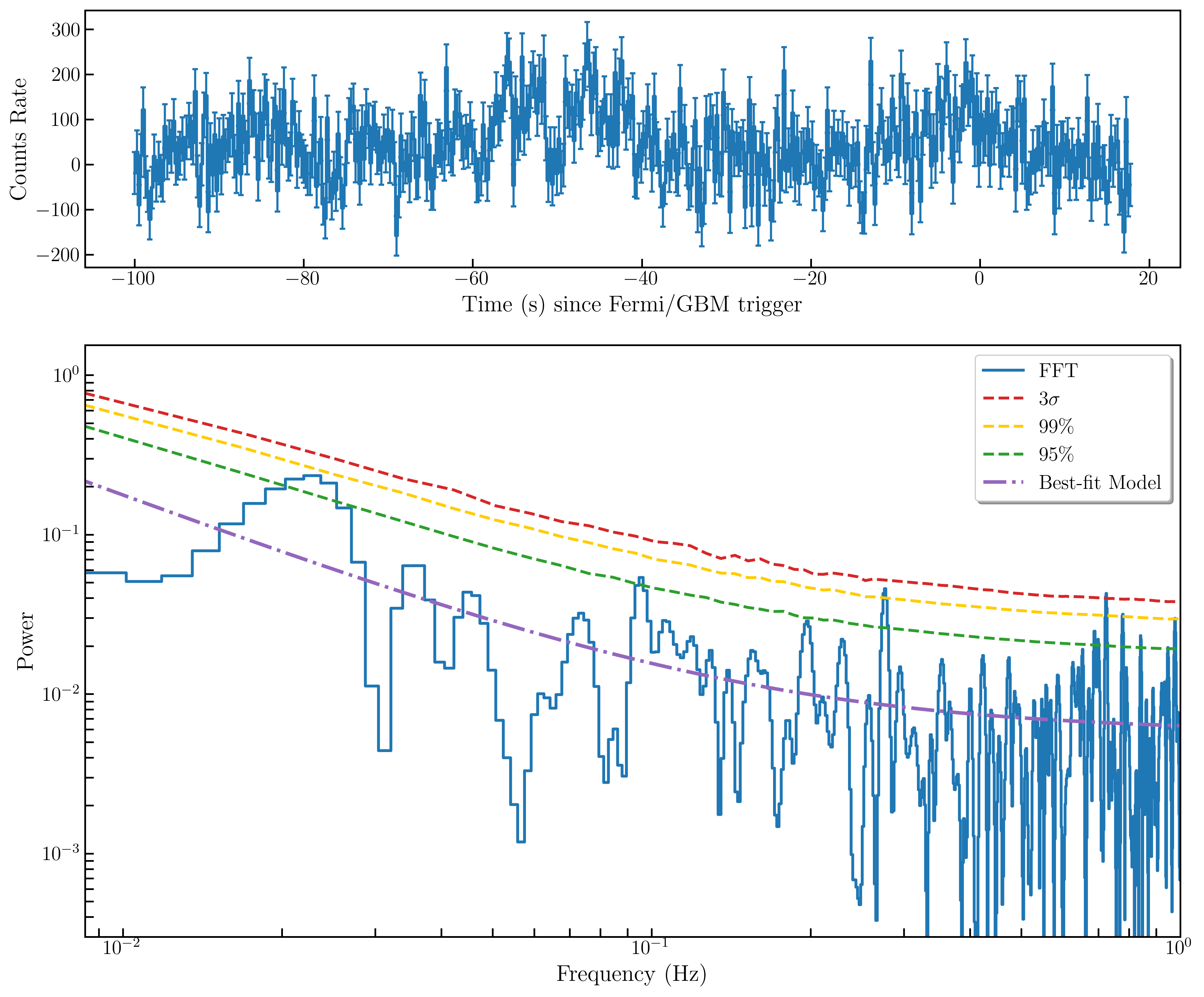}
    {}
    \includegraphics[width=0.75\textwidth]{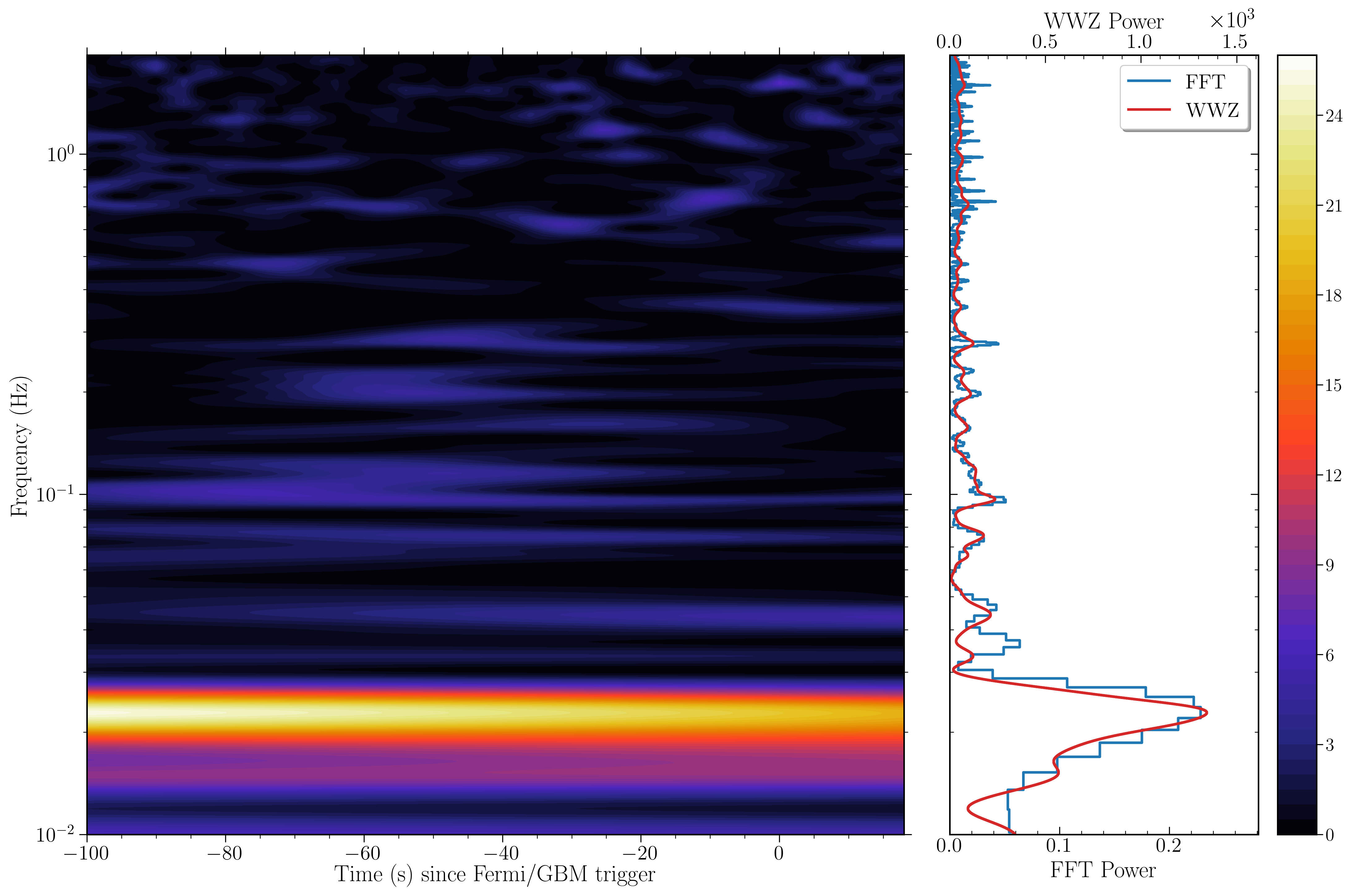}
    \caption{Analysis of GRB 250702D lightcuvre. Top panel: the 256 ms binned lightcurve of GRB 250702D (nb) in the energy band of 8-1000 keV. Middle panel: the PSD of FFT. The dash-dotted line shows the best-fit noise model, with green (95\%), yellow (99\%), and red ($3\sigma$) dashed lines showing the confidence intervals. Bottom-left panel: WWZ power. Bottom-right panel: WWZ time-integrated PSD and FFT PSD.}
        \label{fig:D}
\end{figure}

\begin{figure}
    \centering
    \includegraphics[width=0.75\textwidth]{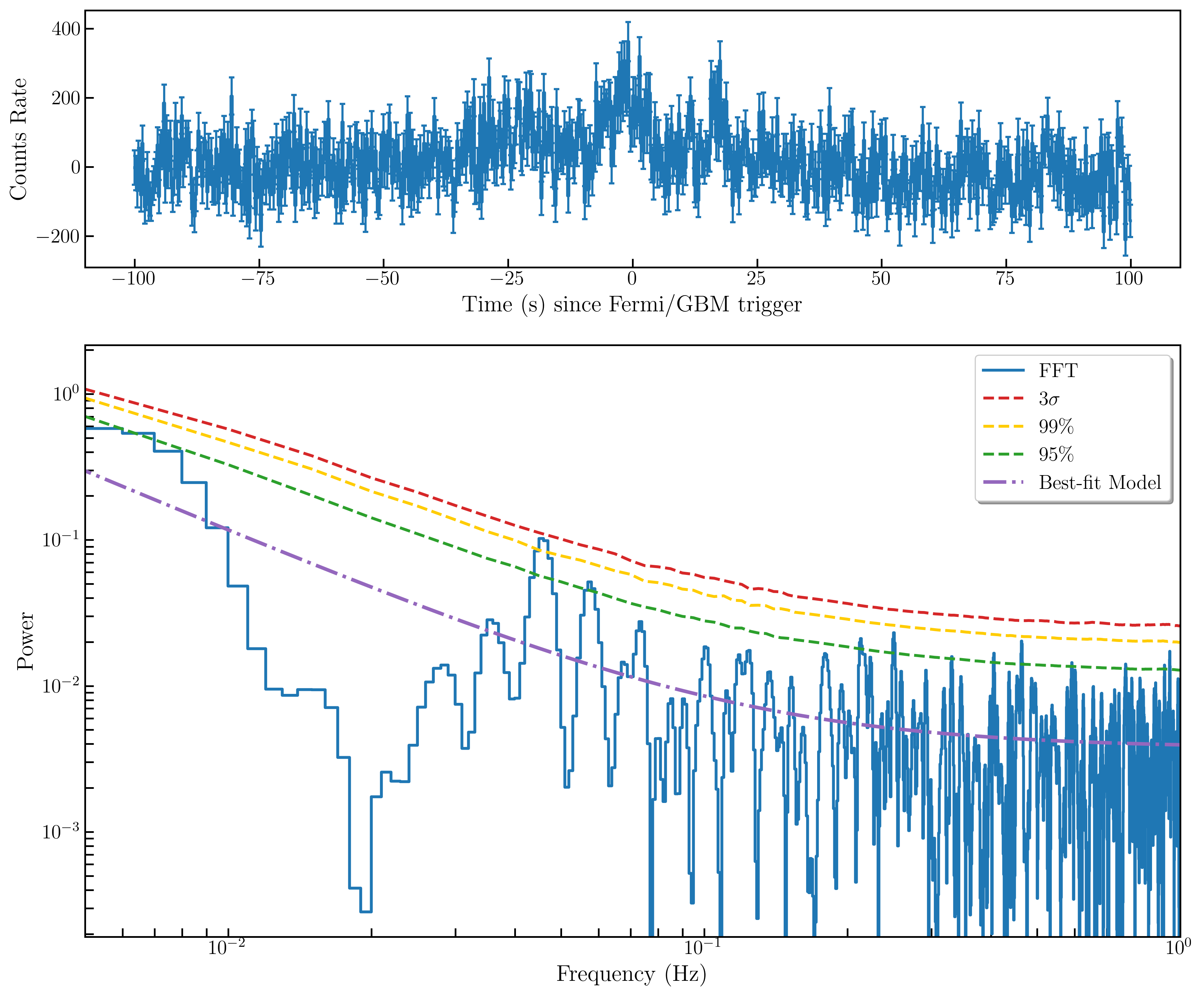}
    {}
     \includegraphics[width=0.75\textwidth]{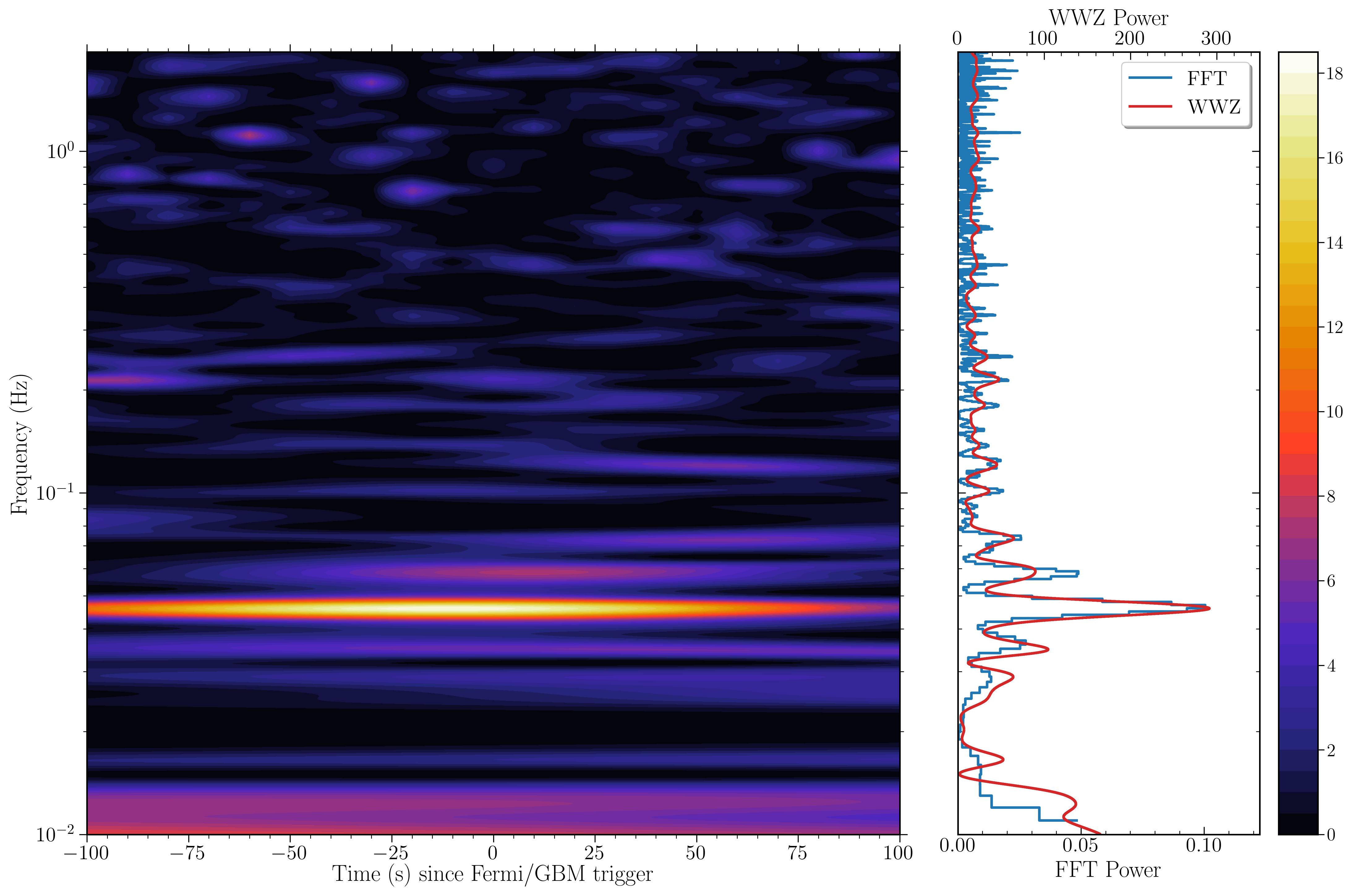}
    \caption{Analysis of GRB 250702B lightcuvre. Top panel: the 256 ms binned lightcurve of GRB 250702B (nb) in the energy band of 8-1000 keV. Middle panel: the PSD of FFT. The dash-dotted line shows the best-fit noise model, with green (95\%), yellow (99\%), and red ($3\sigma$) dashed lines showing the confidence intervals. Bottom-left panel: WWZ power. Bottom-right panel: WWZ time-integrated PSD and FFT PSD.}
    \label{fig:B}
\end{figure}

\begin{figure}
    \centering
    \includegraphics[width=0.75\textwidth]{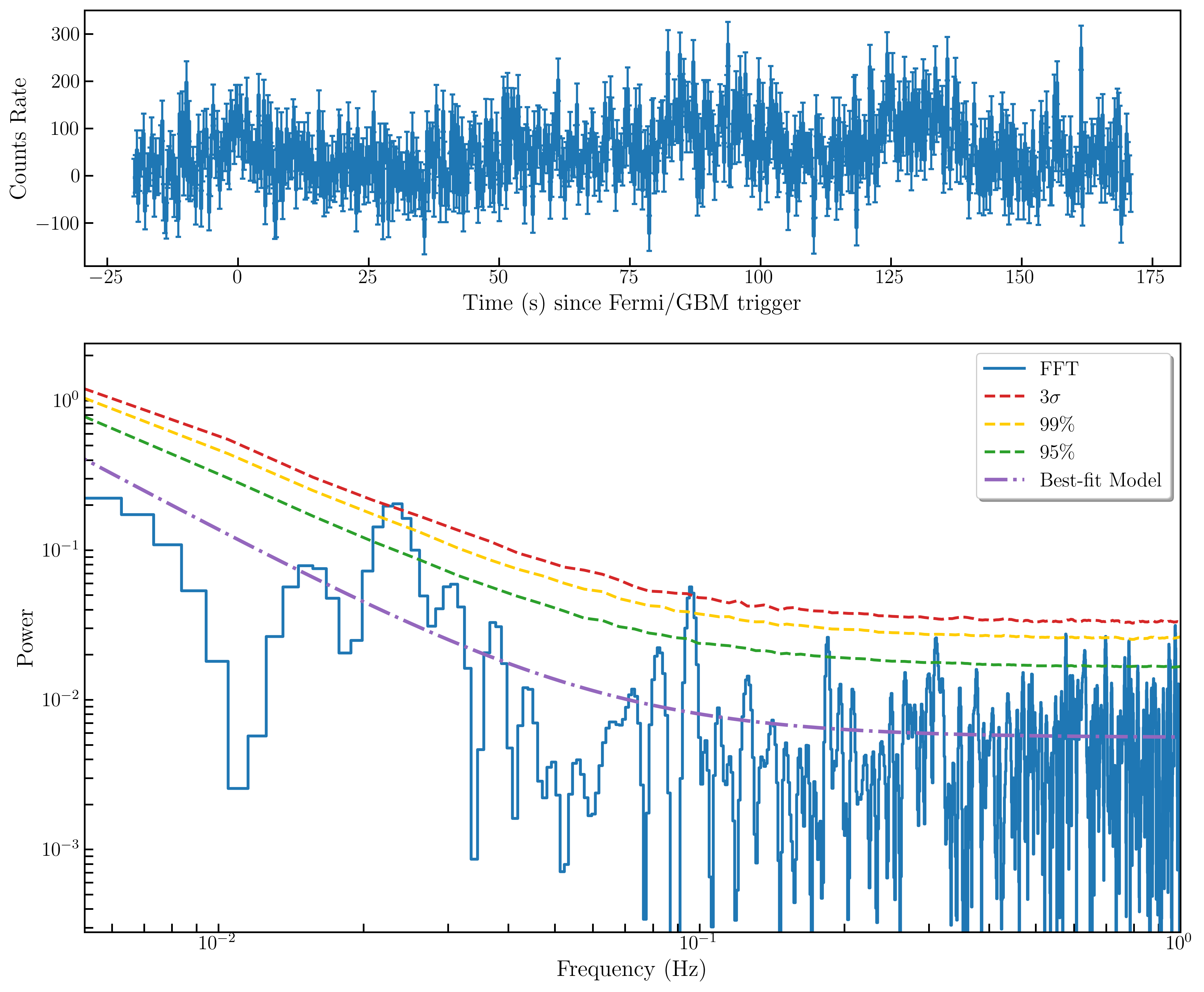}
    {}
    \includegraphics[width=0.75\textwidth]{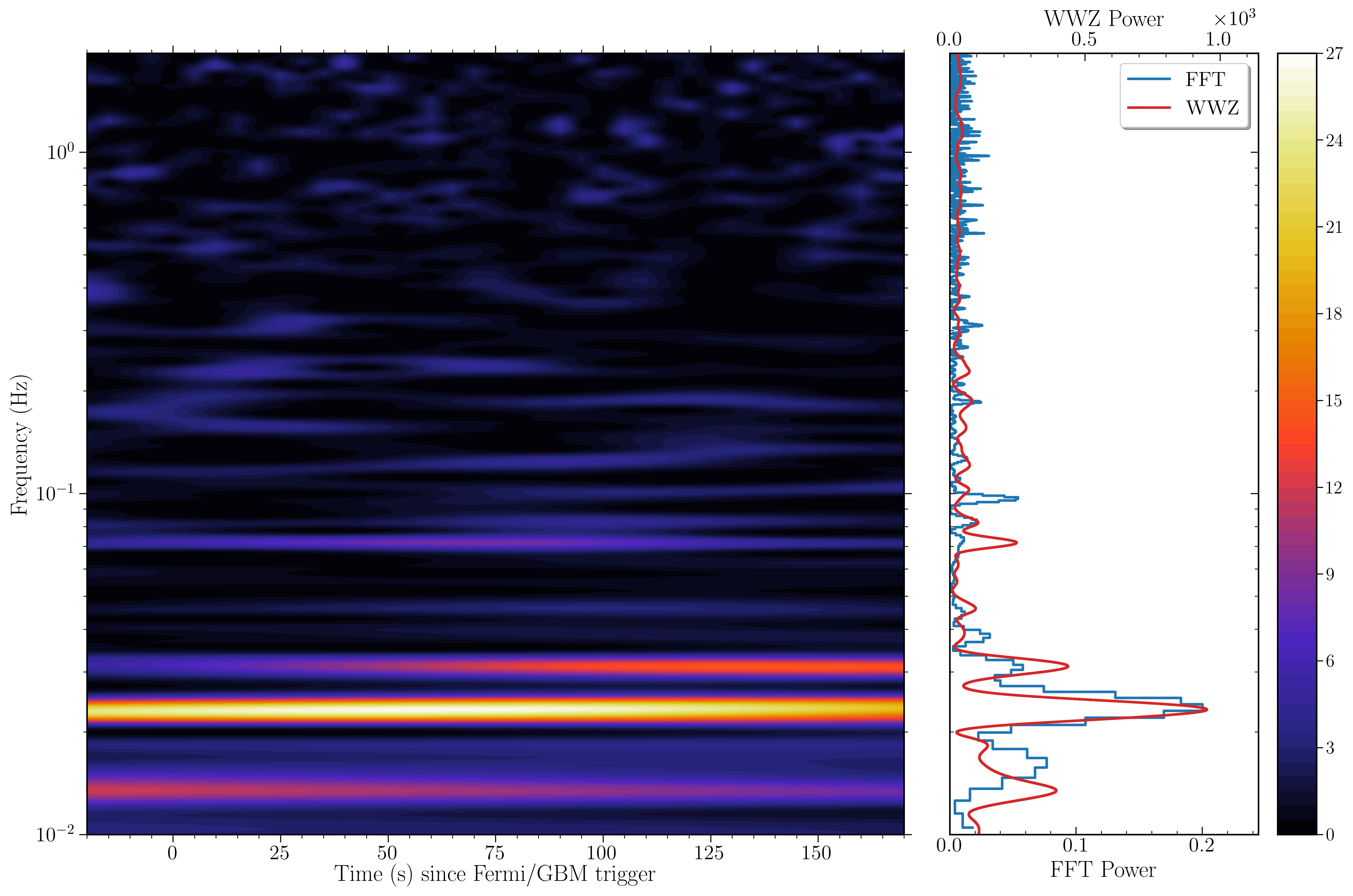}
    \caption{Analysis of GRB 250702E lightcuvre. Top panel: the 256 ms binned lightcurve of GRB 250702E (n8) in the energy band of 8-1000 keV. Middle panel: the PSD of FFT. The dash-dotted line shows the best-fit noise model, with green (95\%), yellow (99\%), and red ($3\sigma$) dashed lines showing the confidence intervals. Bottom-left panel: WWZ power. Bottom-right panel: WWZ time-integrated PSD and FFT PSD. Besides the QPO signal at 0.024 Hz, another possible QPO signal at 0.096 Hz that is exceeding 3$\sigma$ confidence level is noted here. However, this QPO signal is only detected in n8 detector, while it is not shown in the lightcurves from other detectors.  Thus, we are not sure for this identification as a possible QPO signal.}
        \label{fig:E}
\end{figure}

We use FFT and WWZ methods to examine the temporal properties of GRB 250702DBE/EP250702a.
We perform QPO analysis to the lightcurves of this event. 
The lightcurves of GRB 250702DBE/EP250702a are presented in the top panels of Figure \ref{fig:D}-\ref{fig:E}, respectively, with the PSD shown in each bottom-left panel and the WWZ power spectrum displayed in each bottom-right panel. We do not detect any reliable QPO signals in the lightcurve of GRB 250702D.
The PSD of GRB 250702B reveals a QPO signal at 
0.046 Hz corresponding to a period of 21.7 s with almost 3$\sigma$ significance. 
This QPO feature is consistently detected in the WWZ power. 
In GRB 250702E, the PSD reveals a distinct QPO signal at 
0.024 Hz that corresponds to a period of 41.7 s, significantly exceeding $3\sigma$ confidence level.
This QPO signal is clearly shown 
in the WWZ power. 
In addition, by this kind of FFT analysis, we present how often the periodogram is large.


It is noted that we focus on the searching of QPO signal hidden in the low-frequency red noise component. While some signals in the high-frequency white noise component exceed $3\sigma$ significance, we do not consider them to be possible QPO signals for two reasons. First, our results indicate that the variance of high-frequency noise in the FFT PSD is non-stationary, whereas our significance estimation method assumes a constant variance. This discrepancy may lead to inaccuracies in the significance estimation at high frequencies. Second, the WWZ analysis of the high-frequency component does not reveal any statistically significant signal in that regime.

The method of FFT presented in Section \ref{data} for estimating signal confidence only provides how often a signal exhibits high power values in a PSD. 
In order to examine how often a QPO signal occurs, we further perform the analysis below. Based on the best-fit noise model, we generate 20,000 simulated lightcurves. Using the PSDs of these lightcurves, we identify those peaks that exceed the significance level 
of $3\sigma$ in each PSD.
For a candidate peak that has already been identified as the statistically significant, we further classify it as a QPO following the two steps.
First step, the quality factor $Q=f/w$ must satisfy $Q>2$, where $ f $ is the central frequency and $w$ is the full width at half maximum (FWHM) obtained by fitting a Lorentzian function. This criterion has been usually proposed by QPO searching in X-ray binaries \citep{Qfactor} and active galactic nuclei \citep{AGN2008Natur.455..369G}. Thus, we calculate each quality factor Q corresponding a peak with the significance level of $3\sigma$ in each PSD.
Second step, we further examine the peaks both above the significance level of $3\sigma$ and $Q>2$ in each PSD.
We understand that a significant QPO in a PSD is not only a large power value at a certain frequency, but also the large value at the certain frequency with smaller power values at frequencies just below and just above that frequency. 
We evaluate FWHM of a peak signal at a frequency. Then, we extend 5 times length of the FWHM on both sides of the signal. Thus, we compute the average power within the extended range. If this average power falls below the 1$\sigma$ confidence level, 
the signal is considered to satisfy the above criterion. 
Thus, we can clearly identify the peaks with the significance level of $3\sigma$ and the peak profile of $Q>2$ in a PSD. Moreover, the large value at a certain frequency with smaller power values at frequencies just below and just above that frequency can be also identified. We perform the above steps by the blind search in the whole red noise region in a PSD.
Conducting such blind search for any signal throughout the red noise region that satisfies the previously defined criteria in all the 20,000 PSDs, 
we obtain that only 4.56\% of the simulated noise PSDs exhibit a signal meeting the criteria for GRB 250702B, and only 2.10\% for GRB 250702E. 
Therefore, we illustrate how often a possible QPO exsits.
These results indicate that the detected QPO signals are unlikely to originate from noise.

We should examine the false probability given by FFT method. We adopt the approach from \cite{FAP1} to estimate the global false alarm probability (FAP) under the red noise model.
Specifically, under the red noise model, the test statistic $\gamma=2\mathrm{P}_{\mathrm{obs}}(f)/\mathrm{P}_{\mathrm{model}}(f)$ follows a $\chi^2$ distribution with 2 degrees of freedom (where $\mathrm{P}_{\mathrm{obs}}(f)$ is the observed power at a given frequency and $\mathrm{P}_{\mathrm{model}}(f)$ is the power from the red noise model). For the distribution of $\gamma$ at a given frequency, the threshold $\gamma_\epsilon$ is the value for which the exceedance probability is $\epsilon$, i.e., $\mathrm{Pr}(\gamma>\gamma_\epsilon)=\mathrm{exp}(-\gamma_\epsilon/2)=\epsilon$ (see Eq. 14 of \cite{FAP1}). The global FAP for $\gamma$ exceeding $\gamma_\epsilon$  across multiple frequencies is then given by $\mathrm{FAP}=1-(1-\mathrm{exp}(-\gamma_\epsilon/2))^N$. The parameter $N$ represents the effective number of independent frequencies, estimated as $N=(f_{\mathrm{max}}-f_{\mathrm{min}})/\delta f=(f_{\mathrm{max}}-f_{\mathrm{min}})T$, where $\delta f=1/T$ is the frequency resolution of the periodogram, $T$ is the total duration of observation, and $f_{\mathrm{max}}$ and $f_{\mathrm{min}}$ define the frequency range. Our evaluation of the global FAP is conducted within the red-noise region of the FFT PSD. For GRB 250702B and GRB 250702E, the effective number of frequencies in this region amounts to 22 and 11, respectively. The single-trial FAPs for these two cases are estimated as 0.40\% and 0.16\%, respectively. We analytically estimate the global FAP using the $\gamma_\epsilon$ corresponding to the QPO peak as the threshold. The global FAP values are approximately 8.34\% for GRB 250702B and 1.75\% for GRB 250702E.
We then employ Monte Carlo simulations that incorporate red noise characteristics to estimate the global FAP. Using the best-fit red noise model, we generate 20,000 simulated lightcurves and compute their corresponding PSD. For each simulated PSD, we identify the most statistically significant peak relative to the noise model and record its significance level (i.e. the peak corresponding to the smallest $\mathrm{exp}(-\gamma_\epsilon/2)$ value in the PSD). We then calculate the proportion of these simulated peaks whose significance exceeds that of the possible QPO signal, thereby estimating the global FAP. This yields global FAP estimates of 9.01\% for GRB 250702B and 1.73\% for GRB 250702E. 

We have three lightcurve segments, named GRB 250702D, GRB 250702B, and GRB 250702E respectively, in the single event of GRB250702DBE/EP250702a. 
We take each lightcurve segment as one case to identify the possible QPO signal.  
It is important to note that the global FAP estimation only accounts for the number of frequency trials but does not include a correction for the multiple cases examined. We therefore apply a correction using the formula $1-(1-P)^N$, where P represents the single-case global FAP and N = 3 denotes the number of cases examined. This procedure yields corrected FAP values that take into account 
the trials factor from multiple cases. After this correction, the global FAP values derived with the method from \cite{FAP1} increase to 22.99\% for GRB 250702B and 5.16\% for GRB 250702E, while the global FAP estimates based on the Monte Carlo method become 24.67\% and 5.10\% for GRB 250702B and GRB 250702E, respectively.

Besides FFT method, we consider autoregressive method to examine the lightcurves of GRB 250702DBE/EP250702a.  
Here, we perform the autoregressive spectral analysis for the ligthcurves. 
The presentation of the examinations and the related results are presented in detail in Appendix \ref{a2}.

We treat the background of each lightcurve using a standard way of GRB data analysis. However, the QPO searching may be affected by the background examination. Furthermore, this event as GRB explosion is not confirmed. Thus, we take the QPO search to the full lightcurves of the event without background subtraction. In general, the results of QPO searching obtained from the background-subtracted lightcurves are consistent with those obtained from the full lightcurves.    

Different lightcuves are produced by different GBM detectors. For a certain observation, one detector has a certain viewing angle pointing to the source.
For the NaI instrument, we also examine the periodic signal for each lightcurve that is produced by the detectors with large viewing angles. In general, we see that the signal-to-noise ratio varies from high to low value as the viewing angle of the detectors changes from small to large number. Thus, the QPO signal is testified from the celestial object but not from the instrument oscillation.     

Periodic analysis might be affected by the binning timescale of the lightcuvre. 
In this paper, we take the lightcurves of GRB 250702DBE/EP250702a with the binning of 256 ms. We also examined the lightcurves with different binning timescales as 8 ms, 16 ms, 64 ms, and 128 ms. We do not find any difference for the QPO signal detection with different binning timescale. In fact, since the QPO signal has the number of 40 or 20 s, the smaller binning timescale of less than 256 ms does not take any effect on the QPO detection.  

We focus on the temporal analysis from the data obtained by NaI instrument, because the released energy of this event mainly falls into the energy below 1 MeV. However, we still examine the observed data using the BGO instrument. In the energy band of 200 keV-40 MeV obtained from the b1 detector, we do not find any significant QPO signals in the lightcurves of GRB 250702DBE/EP250702a. Moreover, in addition to temporal analysis, spectral analysis is also an important supplement to the investigation of the QPO signals shown in GRB 250702DBE/EP250702a. The spectral analysis from the NaI data is performed, and the spectral analysis combined by both the NaI data and the BGO data is also carried out. The energy band in the NaI data is relatively narrow, and a simple power law seems suitable for the spectral fitting. The combination spectra from the NaI and BGO detections can be fitted by a broken power-law, which can be modeled by the Band function \citep{Band1993}. The detailed results of the spectral analysis are presented in Appendix \ref{AB}.    



It should be noted that we analyze the lightcurves from seven detectors with viewing angles less than $90^{\circ}$ rather than using a single detector. The results presented in this paper are based on detectors with relatively small viewing angles among the seven detectors. For GRB 250702B and GRB 250702E, although the PSDs of the lightcurves from the other six detectors all exhibit a peak near the QPO frequency, none of these peaks exceeds the 3$\sigma$ confidence level of the noise. Considering the trials factor from multiple detectors would further increase the FAP, implying a corresponding reduction in the likelihood of the presence of a QPO signal.

\section{Discussion and Conclusions}
\label{discussion}
We detect a possible QPO signal of 0.024 Hz corresponding to a period of 41.7 s in GRB 250702E in the energy band of 8-1000 keV. In this energy band, a possible QPO signal of 0.046 Hz corresponding to a period of 21.7 s is also detected in GRB 250702B. However, we do not detect any reliable QPO signals in GRB 250702D. The confidence level of the possible QPO signals is comprehensively examined.


From the observational results of GRB 250702DBE/EP250702a in the optical band, it has been suggested that the event has an extragalactic origin \citep{2025arXiv250714286L}. 
In this paper, we may consider this event related to jet/outflow that can have a helical structure. 
The QPO signals may come from the helical jet moving in a periodic way or from the jet procession. 
GRB 250702D can be viewed as the initial stage shown in the first segment of the event. As the jet may not be generated yet at that time, there should be no QPO signals. GRB 250702B is the second segment of the event, and the jet appeared at this stage. A marginal QPO signal was detected. Finally, a fully developed jet is shown as in the third segment of the event, which is called GRB 250702E, and a clear QPO signal was detected. The QPO frequency in GRB 250702B is higher than that in GRB 250702E, because the initial jet may have a structure of higher helicity.

Gamma-ray burst (GRB) explosion may produce a strong relativistic jet. Searching for QPO signals in GRB temporal phenomenon has been attempted in recent years. 
\citet{2024ApJ...970....6X} detected a 22 Hz QPO signal in the precursor of GRB 211211A.
\citet{2025arXiv250410153H} performed the temporal analysis of GRB 230307A. Two QPO signals, one at 1.2 Hz and the other at 2.9 Hz, were identified.
GRB 240825A may have 6.4 Hz and 0.7 Hz QPOs shown in the thermal and nonthermal components, respectively \citep{2025arXiv250716538L}. Moreover,
\citet{2021ApJ...921L...1Z} and \citet{2024ApJ...966..209S} searched GRB X-ray afterglows to detect QPO signals. However, we note that all the QPO signals of the GRBs mentioned above show strong diversities and the numbers spread over a wide range. If GRB 250702DBE/EP250702a has a GRB origin, this GRB event is very special among other usual GRBs. This unique GRB was born in a very dense environment, or the GRB jet was surrounded by a very dense cocoon.
The increasing flux of the GRB was shown in the radio band \citep{2025GCN.40985....1B}. This indicates that a complete breakout occurred after the periodic emission in the high-energy band.

\citet{2016ApJ...823..113P} suggested a tidal disruption event (TDE) of star by stellar compact object to produce the GRB with ultra-long duration. This is a kind of micro-TDE event \citep{2021SSRv..217...54Z}. We consider that the micro-TDE activity may be phenomenologically identified as ultra-long GRBs with QPO signals in the high-energy band. This scenario can be further extended. \citet{2025MNRAS.537.1220R} proposed that a disk is formed around an intermediate-mass black hole, and micro-TDE as a dynamical effect may be involved in this system. Sudden bursts may occur as unique transients in this case. This system can be inside a gas-rich global cluster such that the optical emission suffers very strong absorption. It seems that this extended scenario is consistent with the observations of GRB 250702DBE/EP250702a in the high-energy band and in the optical band.

In principle, the QPO signal detected in GRB 250702DBE/EP250702a only indicates a periodic characteristic of this event. 
It can be classified as any kind of object that can produce periodic emission in the high-energy band. We cannot arbitrarily rule out other possibilities for the progenitor of GRB 250702DBE/EP250702a. Long-term monitoring of this event by multiwavelength telescopes is expected in the future.

\begin{acknowledgments}
This work is supported by the National Key R\&D Program of China (2023YFE0101200), Natural Science Foundation of China 12393813, CSST grant CMS-CSST-2025-A07, and the Yunnan Revitalization Talent Support
Program (YunLing Scholar Project).
We thank the Fermi/GBM team for providing the public data used in this analysis.
Our analysis employs the following packages: Matplotlib \citep{matplotlib2007CSE.....9...90H}, NumPy \citep{numpy2020Natur.585..357H}, SciPy \citep{scipy2020NatMe..17..261V}, Stingray \citep{stingray2019ApJ...881...39H}, and WWZ \citep{wwzm_emre_aydin_2017_375648}. We extract lightcurves and energy spectra using Fermi GBM Data Tools \citep{GbmDataTools}. We perform spectral analysis using XSPEC v12.13.1 \citep{Xspec}.
\end{acknowledgments}

\appendix

\section{Examination from Autoregressive Spectral Analysis}
\label{a2}
\begin{figure}
    \centering
    \includegraphics[width=0.45\textwidth]{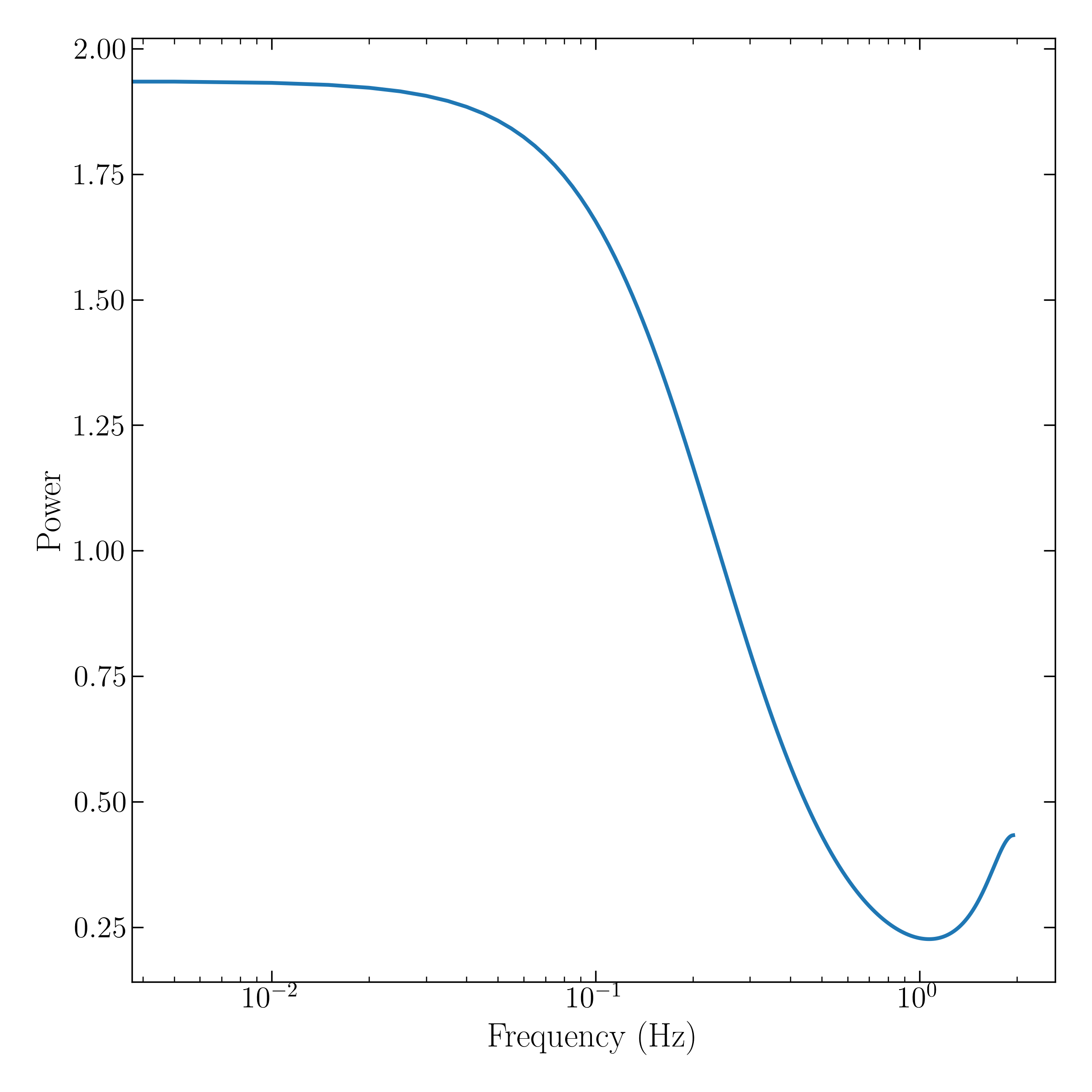}
    {}
    \includegraphics[width=0.45\textwidth]{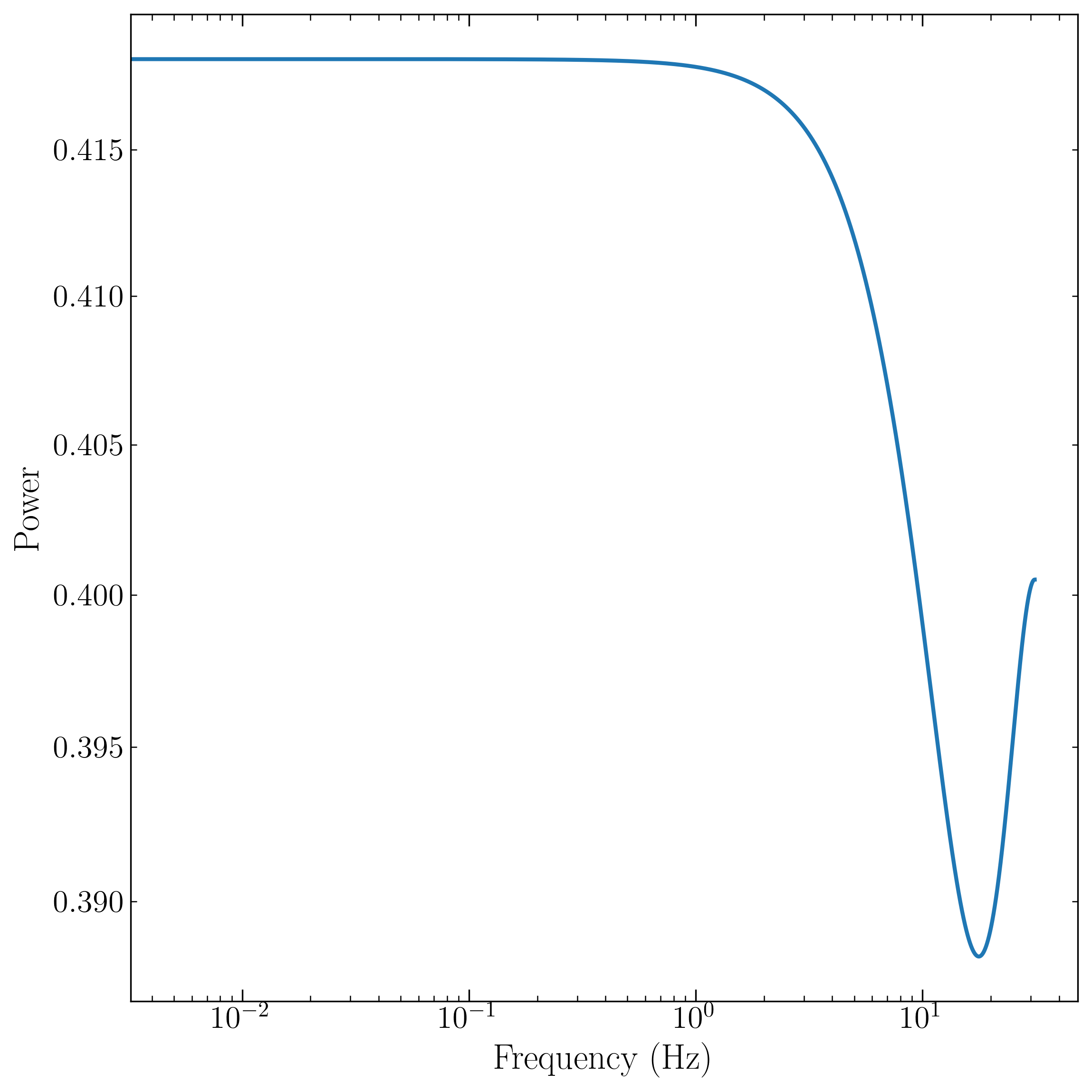}
    {}
    \includegraphics[width=0.45\textwidth]{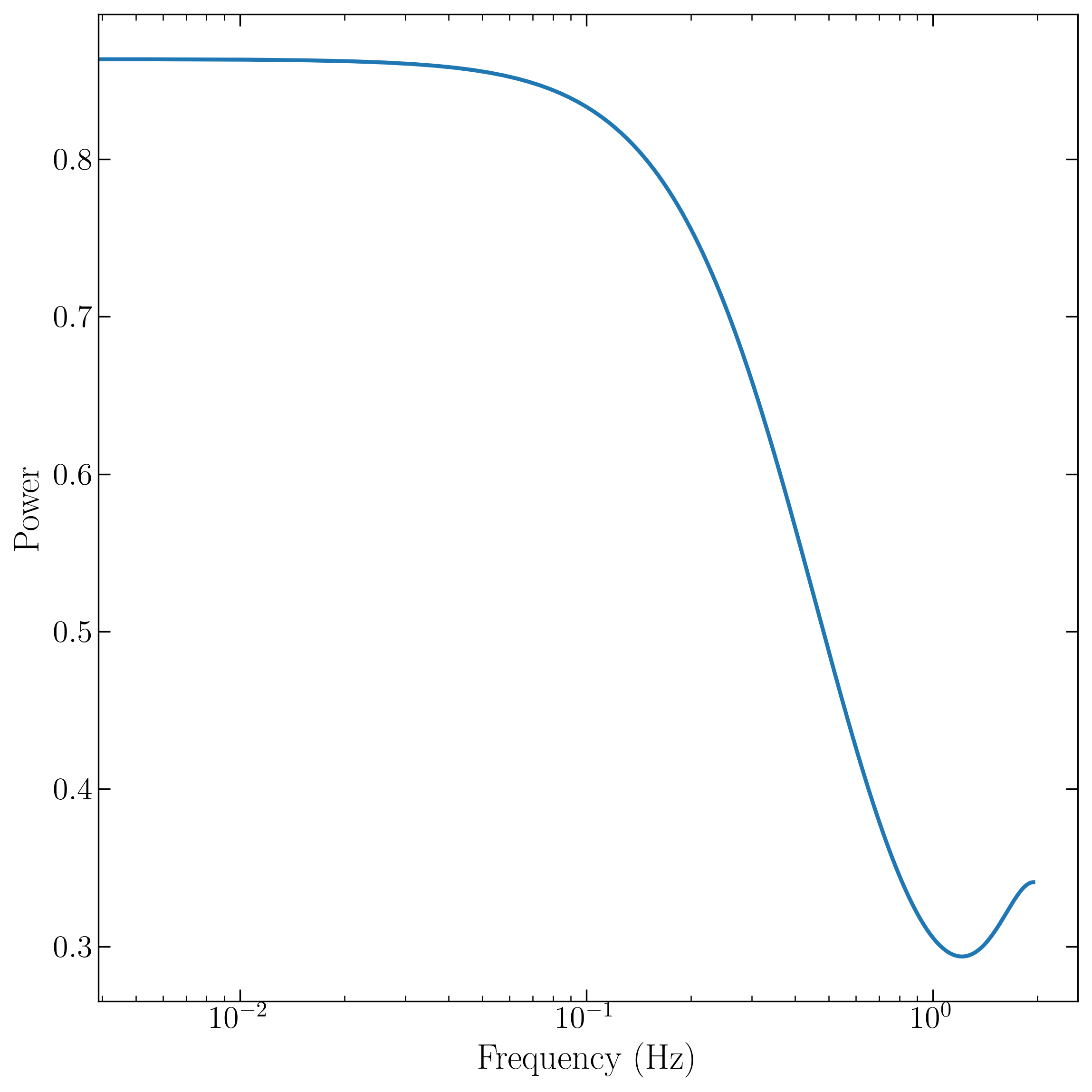}
    {}
    \includegraphics[width=0.45\textwidth]{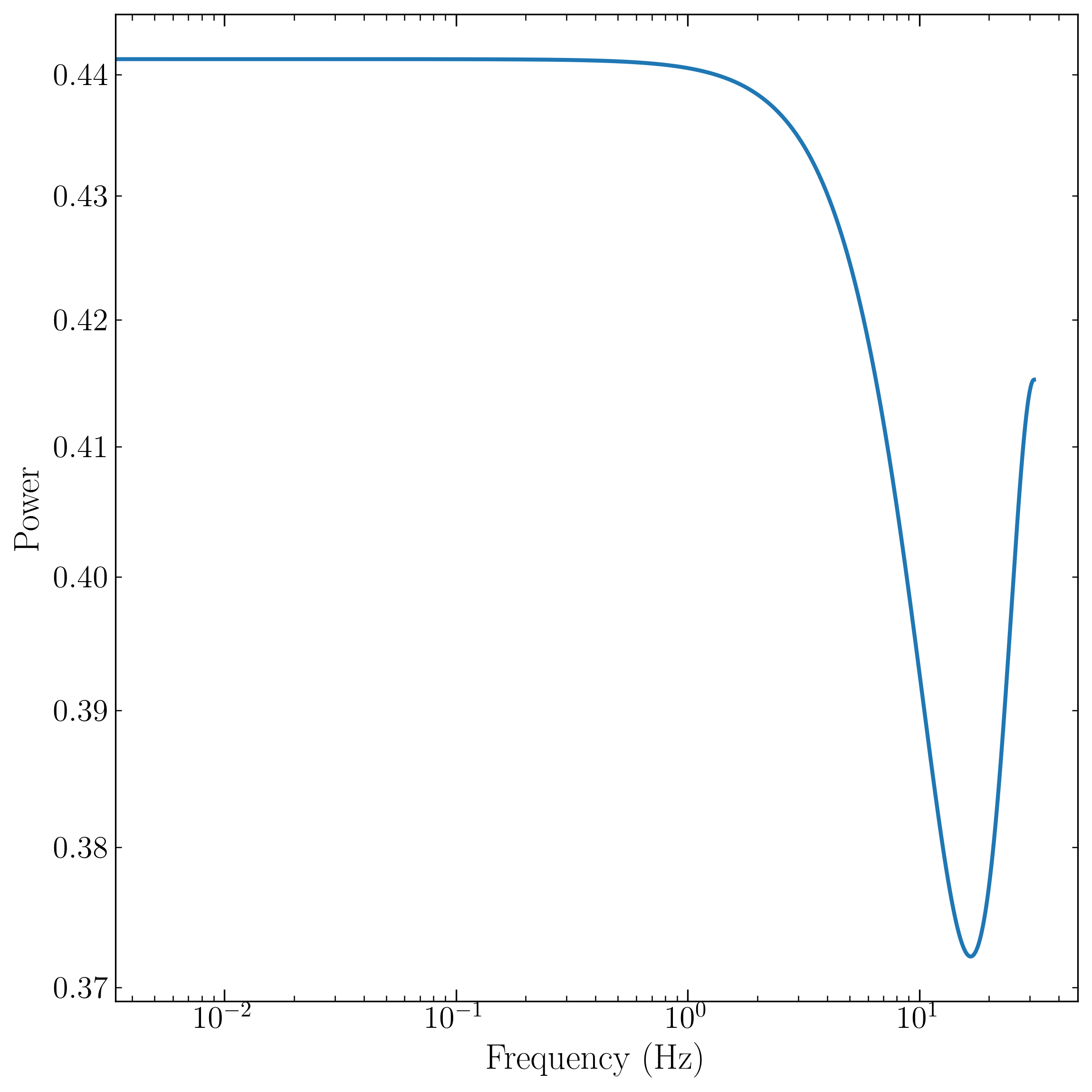}
    \caption{Top panels: PSDs obtained from AR(2) spectral analysis to GRB 250702B lightcurves with 256ms (left) and 16ms (right) time bins. Bottom panels: PSDs obtained from AR(2) spectral analysis to GRB 250702E lightcurves with 256ms (left) and 16ms (right) time bins.}
    \label{fig: AR PSD}
\end{figure}

The lightcurve of GRB 250702DBE/EP250702a is quite random. We employ autoregressive (AR) models for the lightcurve analysis. Specifically, the second-order autoregressive model (AR(2)), defined as $x_t=\phi_1x_{t-1}+\phi_2x_{t-2}+\epsilon_t$, may 
produce a power spectrum with a broad peak structure, when the lightcurve has a characteristic of a random process. Notably, when the condition $\phi_1^2+4\phi_2<0$ is satisfied, the power spectrum of the AR(2) model exhibits a peak. 
We use the AR(2) spetral analysis to the lightcurves of GRB 250702B and GRB 250702E. 
For the 256ms-binned lightcurves, the estimated parameters are $\phi_1=0.253$ and $\phi_2=0.293$ for GRB 250702B, and $\phi_1=0.201$ and $\phi_2=0.119$ for GRB 250702E. The analysis of 16ms-binned lightcurves yield  $\phi_1=0.011$ and $\phi_2=0.013$ for GRB 250702B, and $\phi_1=0.015$ and $\phi_2=0.035$ for GRB 250702E.
In all cases, we obtain $\phi_1^2+4\phi_2>0$. It means that the condition for generating a peak in the power spectrum is not satisfied. The corresponding PSDs of the AR(2) models are presented in Figure \ref{fig: AR PSD}.
The absence of a wide peak in PSDs is clearly shown. 
As we mentioned above, to check the effect of the lightcurve binning on the power spectrum, we take two cases. One is the case of 256 ms binning, the other is the case of 16 ms binning.
It seems that the lightcurve binning does not take effects on the searching of a peak shown in the power spectrum.

\begin{figure}
    \centering
    \includegraphics[width=0.6\textwidth]{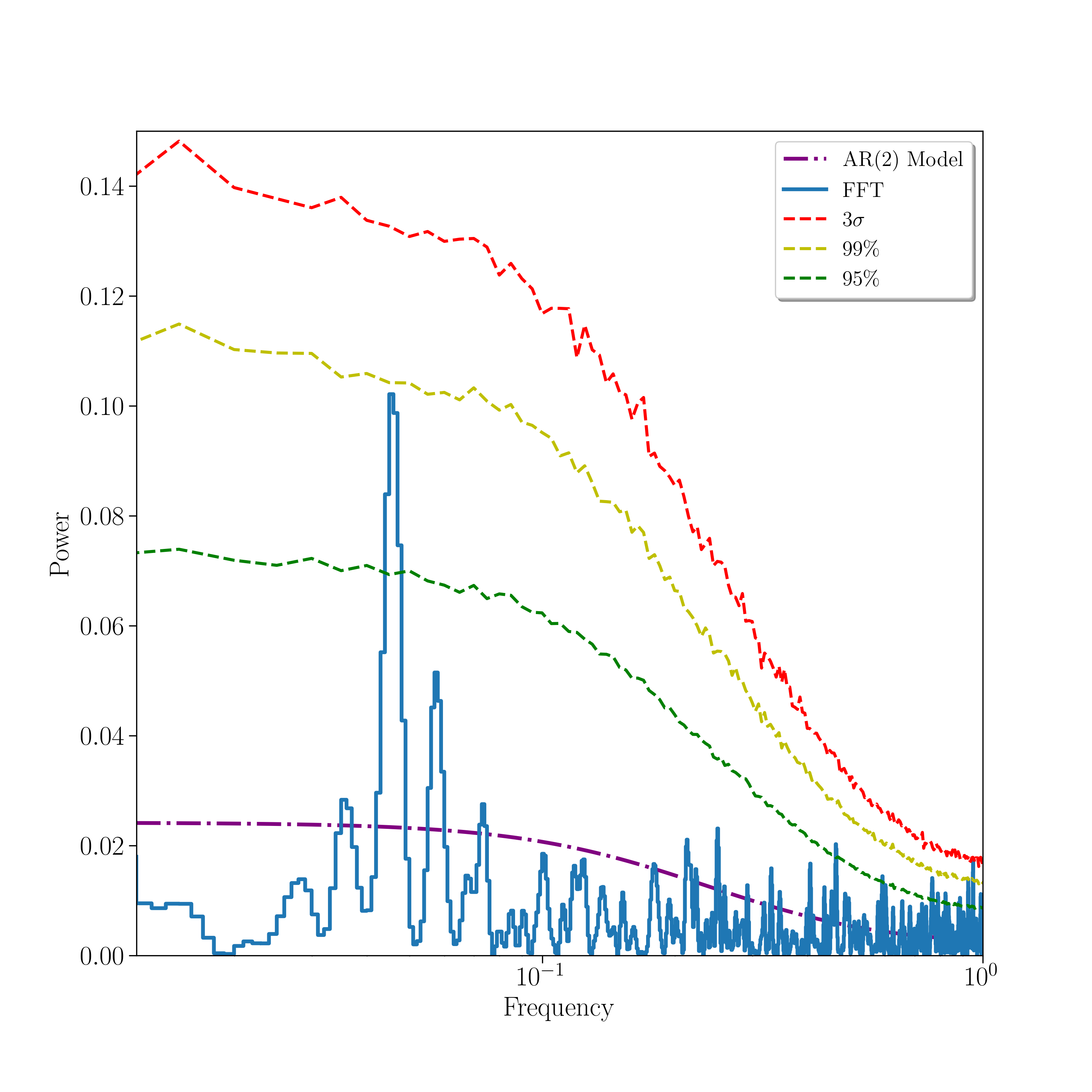}
    {}
    \includegraphics[width=0.6\textwidth]{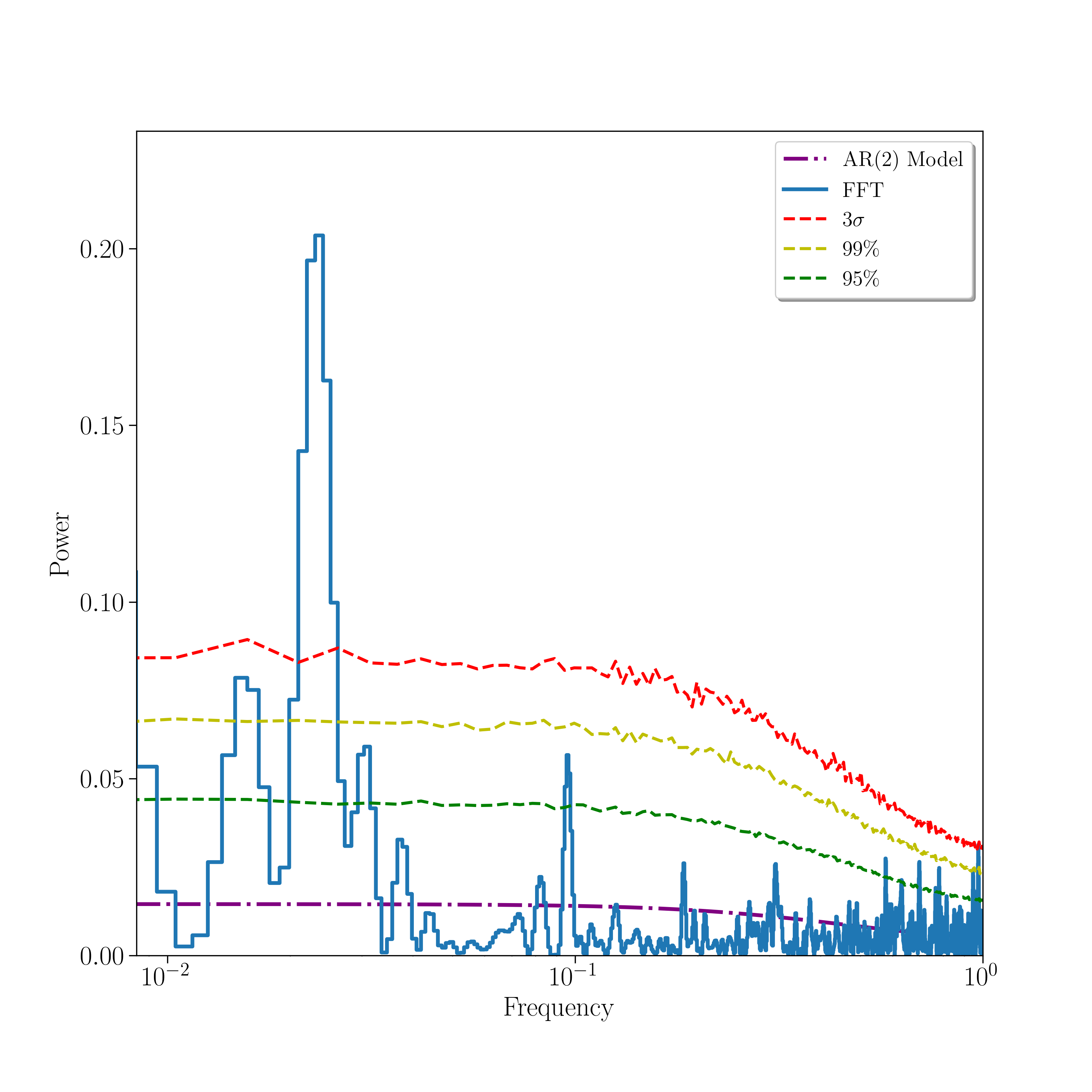}
    \caption{Top panel: PSD comparison between FFT (blue solid line) and AR(2) model (purple dash-dotted line) for GRB 250702B. Dashed lines show AR(2) confidence levels: green (95\%), yellow (99\%), red (3$\sigma$). Bottom panel: Same as top panel but for GRB 250702E. To clearly show the possible QPO signals compared to the AR(2) model, the low frequency ($<10^{-2}$ Hz) region is not displayed in this figure. 
    }
    \label{fig: AR sig}
\end{figure}

We may consider to use the AR(2) spectral results to assess the significance of the QPO signals in GRB 250702B and GRB 250702E. 
We use the spectral results of AR(2) to compare the results obtained from FFT. 
We generate 20,000 time series based on the AR(2) model obtained from the 256ms-binned lightcurve analysis, establishing confidence levels for the PSD of model. Our analysis reveals that the QPO in GRB 250702B approaches the 99\% confidence level of the AR(2) model, while the QPO in GRB 250702E surpasses the 3$\sigma$ confidence level (see Figure \ref{fig: AR sig}).

The results given by AR(2) spectral analysis suggest that the detected QPO signals in Section 3 may unlikely arise 
from the random process in the lightcurves.

\section{Spectral Analysis}
\label{AB}
In the energy band of 8-1000 keV, we extract the energy spectra of GRB 250702D, GRB 250702B, and GRB 250702E 
using Fermi GBM data tools, focusing on the time intervals where the background is excluded. 
For GRB 250702D, we select the spectrum from the time interval [T$0$-100s, T$0$+18s].
For GRB 250702B, we select the spectrum from the time interval [T$'_0$-100s , T$'_0$+100s]. For GRB 250702E, we select the spectrum from the time interval [T$''_0$-20s , T$''_0$+171s].

We perform spectral fitting 
using four models: power-law model, cutoff power-law model, power-law plus blackbody model, and the Band function model. The power-law model is described by
\begin{equation}
N(E) = AE^{-\Gamma},
\end{equation}
where $\Gamma$ represents the dimensionless photon index, and $A$ denotes the normalization factor. 
The cutoff power-law model is defined as
\begin{equation}
N(E) = A E^{-\Gamma} \exp(-E / E_{\rm cut}),
\end{equation}
where $\Gamma$ represents the dimensionless photon index, $E_{\rm cut}$ is the cutoff energy, and $A$ denotes the normalization factor. 
The blackbody model is given by
\begin{equation}
N(E) = \frac{A \times 8.0525 E^{2} dE}{(kT)^{4} [\exp(E/kT) - 1]},
\end{equation}
where $kT$ represents the temperature in keV, and $A$ is the normalization factor. 
The Band function \citep{Band1993} provides the standard spectral model for GRBs, defined as
\begin{equation}
N(E) = \begin{cases}
A(E/100 \rm kev)^{\alpha}\exp(-E/E_c) & \text{for } E < E_c(\alpha - \beta) \\
A[(\alpha-\beta)E_c/100\rm kev]^{(\alpha-\beta)}(E/100\rm kev)^{\beta}\exp(\beta-\alpha) & \text{for } E \geq E_c(\alpha - \beta),
\end{cases}
\end{equation}
where $\alpha$ represents the low-energy power-law index, $\beta$ denotes the high-energy power-law index, $A$ gives the normalization factor, and $E_c$ is the characteristic energy in keV. The peak energy is related to $E_c$ as $(2+\alpha)E_c$.

We summarize the results of the spectral fitting in Table \ref{tab:spec_results}. Although we tested the data by the four models, we prefer that a simple power-law model can roughly describe
the spectral property of GRB 250702DBE/EP250702a.

\begin{figure}
    \centering
    \includegraphics[width=0.55\textwidth]{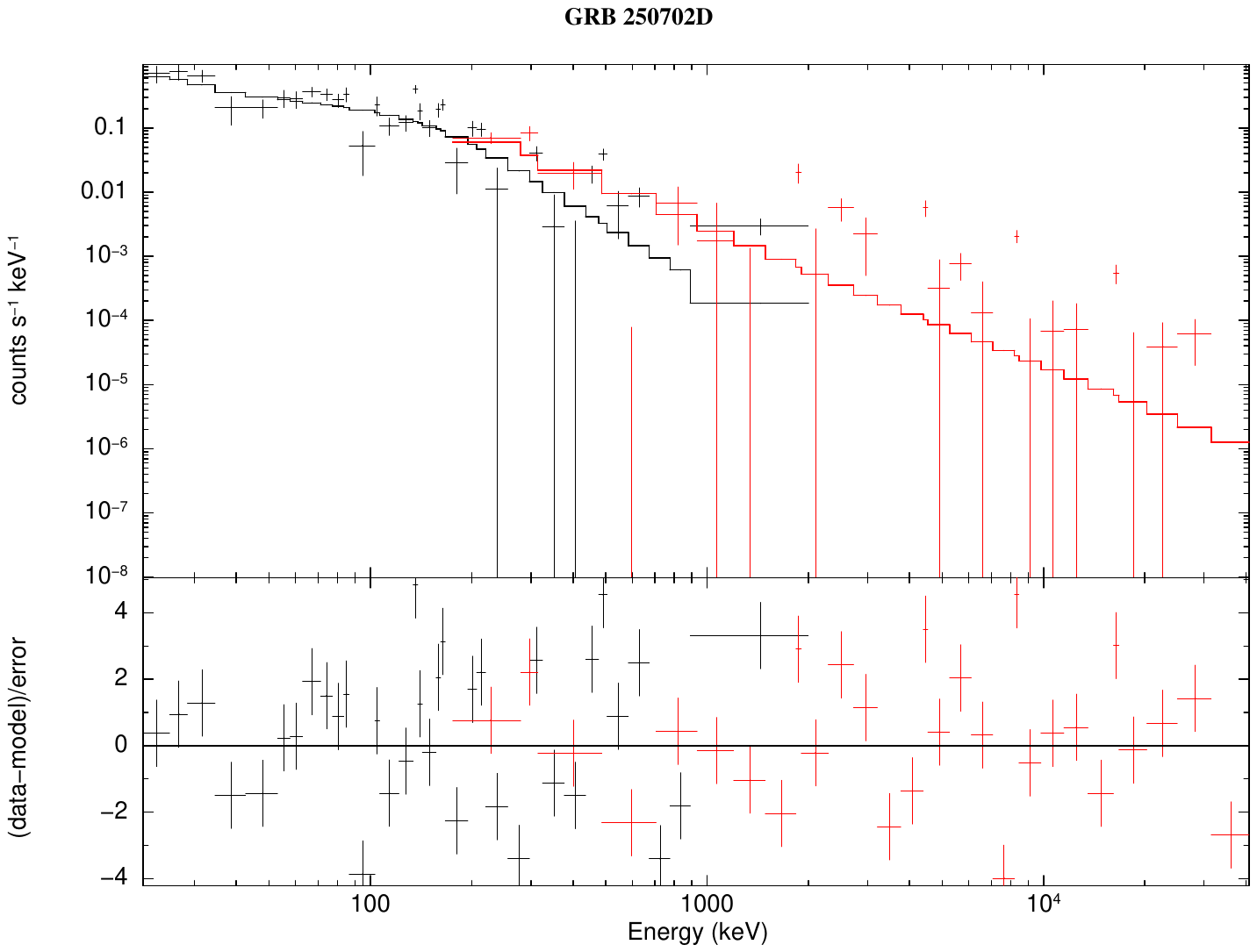}
    {}
    \includegraphics[width=0.55\textwidth]{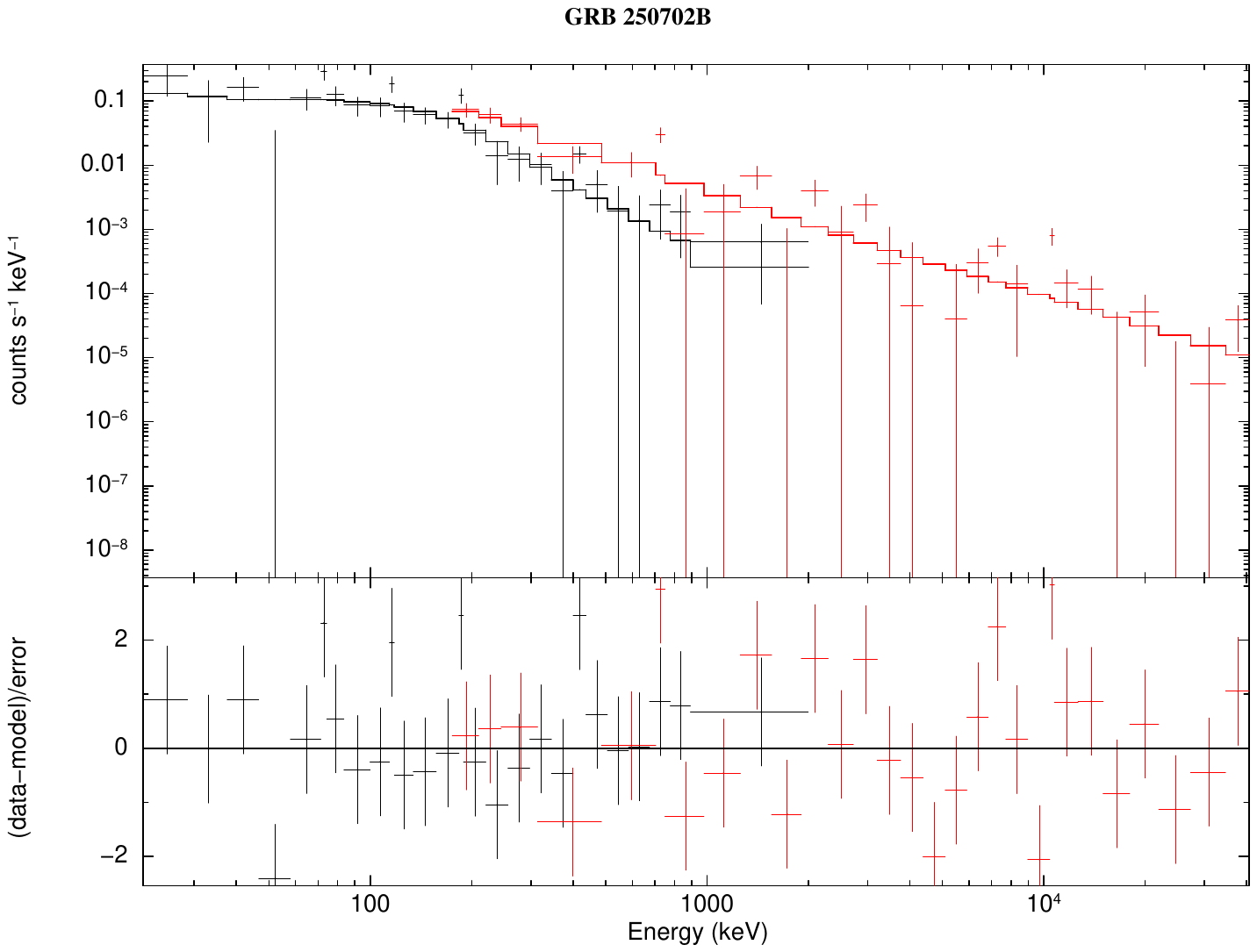}
    {}
    \includegraphics[width=0.55\textwidth]{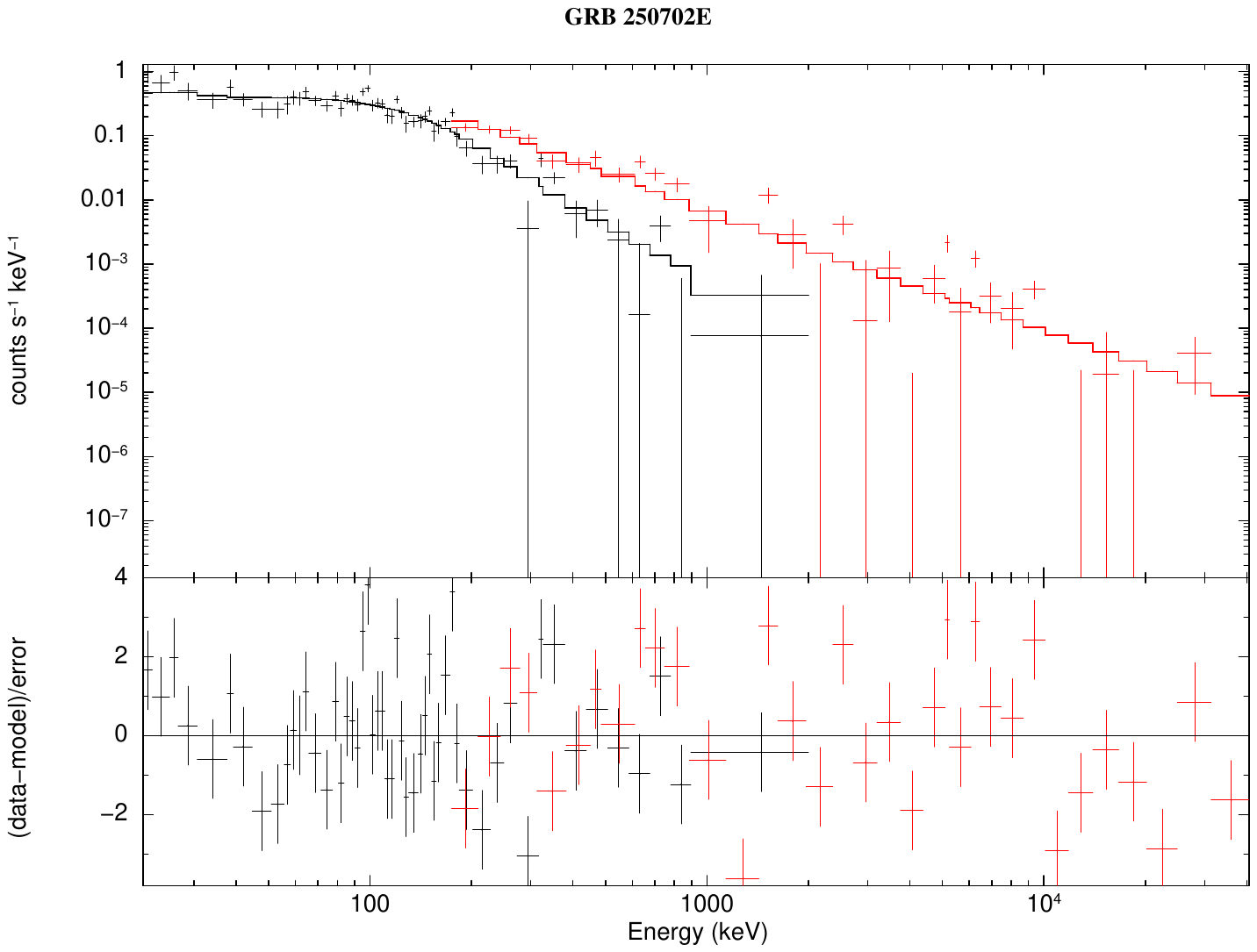}
    \caption{The spectrum of GRB 250702DBE fitted with the Band function. The black data points are from the NaI (nb) detector, while the red data points are from the BGO (b1) detector. The solid stepped line shows the Band function fitting.}
    \label{fig: spec}
\end{figure}

We combine the data in the energy band of 8-1000 keV obtained by NaI instrument and those in the energy band of 200 kev-40 MeV obtained by BGO instrument. 
Thus, the spectra have a wide energy range.
We use the Band function to fit the spectra. The results are shown in Table \ref{tab:spec_results}. It seems that the spectral data fitting in the total energy band from 8 keV to 40 MeV by the Band function provide reasonable results, if GRB250702DBE/EP250702a is assumed to have GRB origin. 
In fact, we can use a broken power-law modeled by the Band function to describe the spectral properties of GRB 250702DBE/EP250702a in general. 
The spectral data and the fitting are shown in Figure \ref{fig: spec}. 

\clearpage

\begin{deluxetable}{lcccc}
\tablecaption{Spectral Fitting Results Comparison. The first 4 models are used to fit the data in the energy band of 8-1000 keV. 
The last model is used to fit the data in the energy band of 8 keV-40 MeV.
\label{tab:spec_results}}
\tablehead{
\colhead{Model/Parameter} &  \colhead{GRB 250702D} & \colhead{GRB 250702B} & \colhead{GRB 250702E} \\
\colhead{} & \colhead{} & \colhead{}
}
\startdata
\textbf{Powerlaw Model} & & \\
\phantom{--} Photon index $\Gamma$& $1.26 \pm 0.06$ & $1.04 \pm 0.10$ & $1.25 \pm 0.03$ \\
\phantom{--} $\chi^2$/dof &5.46& 1.89 & 3.43 \\
\textbf{Cut-off Powerlaw Model} & & \\
\phantom{--} Photon index $\Gamma$& $1.15 \pm 0.17$ & $0.74 \pm 0.33$ & $0.11 \pm 0.19$ \\
\phantom{--} Cut off energy (kev)& $2051.54 \pm 3515.76$ & $1329.67 \pm 1962.21$ & $128.17 \pm 25.86$ \\
\phantom{--} $\chi^2$/dof &5.51& 1.90 & 2.88 \\
\textbf{Powerlaw+Blackbody Model} & & \\
\phantom{--} Blackbody temperature $kT$ (keV) & $36.68 \pm 15.65$ & $44.11 \pm 22.35$& $38.65 \pm 3.22$ \\
\phantom{--} Photon index $\Gamma$ & $1.25 \pm 0.12$ & $0.93 \pm 0.25$ & $1.16 \pm 0.12$\\
\phantom{--} $\chi^2$/dof &5.53 &1.91 & 2.80 \\
\textbf{Band Model} & & \\
\phantom{--} Low-energy powerlaw index $\alpha$ & $-0.89 \pm 0.50$ & $-0.26 \pm 1.23$ & $0.37 \pm 0.37$ \\
\phantom{--} High-energy powerlaw index $\beta$ & $-1.45\pm 0.31$& $-1.25 \pm 0.35$ & $-2.00 \pm 0.26$\\
\phantom{--} $E_{\rm c}$ (keV) & $322.78 \pm 680.94$& $194.18 \pm 492.75$ & $75.64 \pm 23.73$\\
\phantom{--} $\chi^2$/dof &5.55 &1.91 & 2.84 \\
\hline
\textbf{Band Model} & & \\
\phantom{--} Low-energy powerlaw index $\alpha$ & $-0.89 \pm 0.24$ & $-0.17 \pm 0.69$ & $0.33 \pm 0.35$ \\
\phantom{--} High-energy powerlaw index $\beta$ & $-2.14\pm 0.68$& $-1.62 \pm 0.12$ & $-1.85 \pm 0.08$\\
\phantom{--} Characteristic energy $E_{\rm c}$ (keV) & $403.43 \pm 316.96$& $211.55 \pm 219.41$ & $75.69 \pm 22.32$\\
\phantom{--} $\chi^2$/dof & 4.72 & 2.15 & 2.72  \\
\enddata
\end{deluxetable}

\clearpage

\clearpage

\clearpage

\clearpage


\clearpage

\clearpage


\clearpage



\clearpage

\clearpage

\bibliography{sample701}{}
\bibliographystyle{aasjournalv7}



\end{document}